\documentclass[conference]{IEEEtran}
\usepackage{amsmath}
\usepackage{amsmath,amssymb}
\usepackage{bm}
\usepackage{url}
\usepackage{ascmac}

\usepackage{graphicx}
\usepackage{color}

\IEEEsettopmargin{t}{0.8in}

\usepackage{comment}

\title{Deep Learning-Aided 
Trainable Projected Gradient Decoding for LDPC Codes}

\author{%
  \IEEEauthorblockN{
  		Tadashi Wadayama
  		and Satoshi Takabe}
  \IEEEauthorblockA{\IEEEauthorrefmark{1}%
		Nagoya Institute of Technology,
		Gokiso, Nagoya, Aichi, 466-8555, Japan,\\
 		\{wadayama, s\_takabe\}@nitech.ac.jp}
}

\begin{document}
%
\maketitle

\begin{abstract}
We present a novel optimization-based decoding
algorithm for LDPC codes that is suitable  for hardware architectures specialized to feed-forward neural networks.
The algorithm is based on the projected gradient descent algorithm with 
a penalty function for solving a non-convex minimization problem.
The proposed algorithm has several internal parameters
such as step size parameters, a softness parameter,  and the penalty coefficients. 
We use a standard tool set of deep learning, i.e., back propagation and 
stochastic gradient descent (SGD) type algorithms,  to optimize these parameters.
Several numerical experiments show that the proposed algorithm outperforms 
the belief propagation decoding in some cases.
\end{abstract}

\section{Introduction}

Low-density parity-check (LDPC) codes have been adopted 
in numerous practical communication and storage systems,
e.g., digital satellite broadcasting, wireless mobile communications,
hard disks and flash memories, and the 5G standard
as a key component to increase the reliability of the information exchange.
Since the combination of LDPC codes and belief propagation (BP) decoding 
is capacity approaching (or achieving in some cases) with practical 
computation complexity, such wide use of the LDPC codes can be seen as 
a natural consequence. If we are allowed to use a code of long length,
this trend would continue in future.

Recently, interest to machine to machine (M2M) communications is increasing 
in the context of internet of things (IoT). In M2M communications, {\em latency} 
of communications has critical importance in order to achieve 
harmonized real time operations of a number of machines. In 5G wireless system,
ultra low latency (up to 1 millisecond) mode will be prepared for M2M communications. 
For error correction of such a system,  a code with short code length, e.g., order of hundred,
is preferable choice in order to reduce the latency.
In such a situation, it is not clear whether BP decoding is the best candidate 
in terms of the decoding performance.

There have been several decoding algorithms that outperform the BP performance.
Ordered Statistics Decoding (OSD) for LDPC codes \cite{Marc} is one of the most known algorithms
in such a category.  Based on the BP decoding, an OSD decoder iteratively produces candidate codewords by a re-encoding process.
An OSD decoding process contains internal processes to find the $k$-largest values among $n$-candidates 
and Gaussian elimination. These operations are not suitable for hardware implementation and difficult to be parallelized.

Another stream of studies trying to find decoding algorithms surpassing the BP performance is
the {\em optimization-based decoding} for LDPC codes.
The origin of the optimization-based coding is the work by Feldman \cite{Feldman} that
presents a linear programming (LP) formulation of decoding for LDPC codes.
Since the Feldman's work, a number of works for optimization-based decoding
have been presented in this decade. One notable work is 
the Alternative Direction Method of Multipliers (ADMM)-based decoding with
the penalty function by Liu and Draper \cite{Draper}. They presented that 
their decoder provides smaller bit error rate (BER) performances than those of BP decoding
with reasonable computational complexity.

In this paper, we present a novel optimization-based decoding
algorithm for LDPC codes that is suitable for hardware architectures specialized to feed-forward neural networks (NN).
This is because the proposed algorithm mostly consists of matrix-vector products and coordinate-wise non-linear map operations.
Based on recent interests in NN-oriented hardware architectures, it is expected that such an architecture will be included in future CODEC-chips.
The algorithm is based on the projected gradient algorithm for solving 
the non-convex minimization problem based on the objective function that is 
a composition of a linear combination of received symbols 
and the penalty functions corresponding to the parity constraints.
The decoding process consists of two steps, i.e., the gradient step and the projection step.
The gradient step is the gradient descent process such that a search point moves to 
the direction of negative gradient of the objective function. The projection step is based on 
a soft projection operator for binary values.

Since the optimization problem we dealt is a non-convex problem, we cannot expect 
the convergence of a gradient descent type algorithm to the global minimum.
A remarkable point of the proposed algorithm is that 
the initial point of a projected gradient process is randomly chosen.
Since a trajectory of the search point depends on the initial point, the output 
of the algorithm may vary for each decoding trial, i.e., 
the output of the algorithm becomes a random variable.
Executing the multiple trials with randomly initial points is a common technique in non-convex optimization
to find the global minimum.
We can expect that multiple decoding trials, called {\em restarting},  can improve the 
decoding performance.

The proposed algorithm has several internal parameters
such as step size parameters for the gradient descent step, 
a softness parameter controlling the softness of the soft projection function, and
penalty coefficients controlling the strength of the penalty term in the objective function.
The appropriate choice of these parameters are of critical importance for the algorithm to work properly.
In this paper, we use a standard tool set of deep learning (DL), i.e., back propagation and 
stochastic gradient method (SGD) type algorithms,  to optimize these parameters.
By unfolding the signal-flow of the proposed algorithm, we can obtain a multilayer signal-flow graph
that is similar to the multilayer neural network. Since all the internal processes are differentiable, 
we can adjust the internal trainable parameters via a training process based on DL techniques.
This approach,  {\em data-driven tuning},  is becoming a versatile technique
especially for signal processing algorithms based on numerical optimization \cite{LISTA} \cite{TISTA}.


\section{Preliminaries}


\subsection{Notation}
In this paper, a vector $\bm{z} \in \mathbb{R}^d$ is regarded as a row vector of dimension $d$.
For $\bm{z} = (z_1, \ldots, z_n) \in \mathbb{R}^n$ and a scalar $c \in \mathbb{R}$,
we will use the following notation:
\begin{equation}
c + \bm{z} := (c+z_1, c + z_2, \ldots, c + z_n)
\end{equation}
for simplicity. For a real-valued function $f: \mathbb{R} \rightarrow \mathbb{R}$,
$f(\bm{z})$ means the coordinate-wise application of $f$ to $\bm{z} = (z_1, \ldots, z_n)$ such that 
$
f(\bm{z}) := (f(z_1), f(z_2), \ldots, f(z_n)).
$
The $i$-th element of $f(\bm{z})$, i.e., $f(z_i)$ is also denoted by $(f(\bm{z}))_i$.
The set of consecutive integers from $1$ to $n$ is denoted by $[n] := \{1,2,\ldots, n\}$.
An $n$-dimensional unit cube  is represented by
\begin{equation}
[0,1]^n := \{(x_1, \ldots, x_n) \in \mathbb{R}^n 
\mid \forall i \in [n], 0 \le x_i \le 1 \}.
\end{equation}
The cardinality or size of a finite set $A$ is represented by $|A|$.
The indicator function $\mathbb{I}[cond]$ takes the value $1$ if the $condition$ is true;
otherwise it takes the value $0$.

\subsection{Channel model}

Let $H$ be an $n \times m$ sparse parity check matrix over $\mathbb{F}_2$ where $n > m$.
The binary LDPC code defined by $H$ is denoted by 
\begin{equation}
C(H) := \{\bm{x} \in  \mathbb{F}_2^n \mid H \bm{x}^T = \bm{0} \}.
\end{equation}
In the following discussion, we consider that $0$ and $1$ in $\mathbb{F}_2$
are embedded in $\mathbb{R}$ as $0$ and $1$, respectively.
The design rate of the code is defined by $\rho := 1 - n/m$.

In this paper, we assume additive white Gaussian noise (AWGN) channels with 
binary phase shift keying (BPSK) signaling.
A transmitter choose a codeword $\bm{c} \in C(H)$ according to the message fed into 
the encoder.
A binary to bipolar mapping is applied to $\bm{c}$ to generate a bipolar codeword
\begin{equation}
\bm{x} := 1 - 2 \bm{c} \in \mathbb{R}^n.
\end{equation}
The bipolar codeword $\bm{x}$ is sent to the AWGN channel and 
the receiver obtains a received word $\bm{y} = \bm{x} + \bm{w}$ where
$\bm{w}$ is an $n$-dimensional Gaussian noise vector with mean zero and 
the variance $\sigma^2/2$.
The signal-to-noise ratio $SNR$ is defined by 
$SNR := 10 \log_{10} (1/(2 \sigma^2 \rho))$ (dB).  
The log likelihood ratio vector corresponding to $\bm{y}$ is 
given by $\bm{\lambda} := 2 \bm{y}/\sigma^2$.
The decoder's task is to estimate the transmitted word from a given 
received word $\bm{y}$ as correct as possible.
The maximum likelihood (ML) estimation can be expressed by a non-convex optimization form:
\begin{equation} \label{MLrule}
	\hat{\bm{x}} := \mbox{argmin}_{\bm{c} \in C(H)} || \bm{y} - (1- 2 \bm{c})||_2^2.
\end{equation}
It is hopeless to solve the problem naively and directly because of its computational complexity.
We need to rely on an approximate algorithm to tackle the problem.

\subsection{Fundamental polytope}

Feldman \cite{Feldman} proposed a continuous relaxation of the ML rule (\ref{MLrule}) based on 
the {\em fundamental polytope}.
The  fundamental polytope is a polytope in $\mathbb{R}^n$ such that
any codeword of $C(H)$ is a vertex of the polytope. In other words,
the feasible region of (\ref{MLrule}), i.e., $C(H)$, is relaxed to the 
fundamental polytope in the Feldman's formulation.
The fundamental polytope is defined by the simple box constraints and
a set of linear inequalities derived from the parity check constraints.
In the following, we will review the definition of the fundamental polytope.

Let $A_i (i \in [m])$ be an index set defined by 
\begin{equation}
	A_i := \{j \in [n] \mid h_{i,j} = 1\}	
\end{equation}
where $h_{i,j}$ denotes the $(i,j)$-element of $H$.
Let $T_i$ be the family of subsets in $A_i$ with odd size, i.e.,
\begin{equation}
O_i := \{S \subset A_i \mid |S|  \mbox{ is odd} \}.
\end{equation}
The {\em parity polytope} ${\cal Q}(H)$ defined 
based on the parity check matrix $H$ is defined by
\begin{equation}
	{\cal Q}(H) := \{ \bm{x} \in \mathbb{R}^n \mid  \bm{x} \mbox{ satisfies the parity constraint (\ref{parityconst})}  \},
\end{equation}
where the parity constraints are given by
\begin{equation}
	\forall i \in [m],\ \forall S \in O_i,\quad  1 + \sum_{t \in S} (x_t -1) 
	- \sum_{t \in A_i \backslash S} x_t \le 0.
	\label{parityconst}
\end{equation}
A parity constraint defines a half-space and the intersection of the half-spaces induced by 
all the parity constraints  is the
parity polytope. These parity constraints introduced by Feldman \cite{Feldman} come from 
the convex hull of the single parity check codes.

The fundamental polytope \cite{Feldman} corresponding to $H$ is 
the intersection of ${\cal Q}(H)$  and $n$-dimensional cube $[0,1]^n$:
\begin{equation}
	{\cal F}(H) := {\cal Q}(H) \cap [0, 1]^n.
\end{equation}
Since the number of the parity constraints for a given $i \in [m]$ is $2^{|A_i| -1}$,  the total number of all the
constraints becomes $n + \sum_{i \in [m]} 2^{|A_i| -1}$.  In the case of LDPC codes, the maximum size of
the row weight, i.e., $\max_{i \in [m]} |A_i|$ is constant to $n$ and thus the total number of 
constraints is 
\begin{equation}
n + \sum_{i \in [m]} 2^{|A_i| -1} = n + \rho n O(1) = O(n).	
\end{equation}

LP decoding \cite{Feldman} is based on the following minimization problem:
\begin{equation} \label{LPdecoding}
	\mbox{minimize}_{\bm{x} \in \mathbb{R}^n}  \bm{\lambda} \bm{x}^T  \mbox{ subject to } \bm{x}  \in {\cal F}(H),
\end{equation}
where $\bm{\lambda} := (\lambda_1, \lambda_2, \ldots, \lambda_n)$.
Since the objective function and all the constraints are linear,  this problem is an LP problem.
The fundamental polytope includes vertices that are not contained in $C(H)$.
This means that LP decoding may produce a non-integral (or factional) solution. It is known
that, if we have an integral solution,  it coincides with the ML estimate. 
This property is called the {\em ML certificate property} of LP decoding.

\section{Trainable Projected Gradient Decoding}

This section describes the proposed decoding algorithm in detail.
Firstly, a basic idea is briefly explained. The following subsections
are devoted to describe the details of the proposed algorithm.

\subsection{Overview}

We start from an unconstrained optimization problem closely related  to the 
LP decoding \cite{Feldman}: 
\begin{equation} \label{penaltyfunction_problem}
	\mbox{minimize}_{\bm{x} \in \{0,1\}^n }  \bm{\lambda} \bm{x}^T +  \beta P(\bm{x}),
\end{equation}
where $P(\bm{x})$ is a penalty function satisfying 
$P(\bm{x}) = 0$ if $\bm{x} \in {\cal Q}(H)$; otherwise $P(\bm{x}) > 0$. The scalar parameter 
$\beta$ called the penalty coefficient that adjusts the strength of the penalty term.
From the ML certificate property, it is clear the solution of (\ref{penaltyfunction_problem})
coincides with the ML estimate if $\beta$ is sufficiently large.
Although the optimization problem in (\ref{penaltyfunction_problem}) is a non-convex problem,
it can be a start point of an numerical optimization algorithm for solving (\ref{MLrule}).

Let 
\begin{equation}
f_\beta(\bm{x}) := \bm{\lambda} \bm{x}^T +  \beta P(\bm{x}),
\end{equation}
which is our objective function to be minimized.
We here use the projected gradient descent algorithm for solving (\ref{penaltyfunction_problem})
in an approximate manner.
The projected gradient descent algorithm consists two steps, 
{\em the gradient step and the projection step}. 
The gradient descent step moves the search point 
along the negative gradient vector of the objective function.
The projection step moves the search point into a feasible 
region. The two steps are alternatively performed

In the gradient step,  a search point is updated in a gradient descent manner, i.e., 
\begin{equation}
\bm{r}_t  := \bm{s}_t - \gamma_t \nabla f_{\beta_t}(\bm{s}_t),
\end{equation}
where $\nabla f_{\beta_t}(\bm{x})$ is the gradient of $f_{\beta_t}(\bm{x})$. The index $t$ represents the iteration index.
A scalar $\gamma_t \in \mathbb{R}$ is the step size parameter. 
If the step size parameter is appropriate,
a search point moves to a new point having a smaller value of the objective function.
The parameter $\beta_t \in \mathbb{R}$ is an iteration-dependent penalty coefficient.

\begin{figure}
\begin{center}
	\includegraphics[scale=0.8]{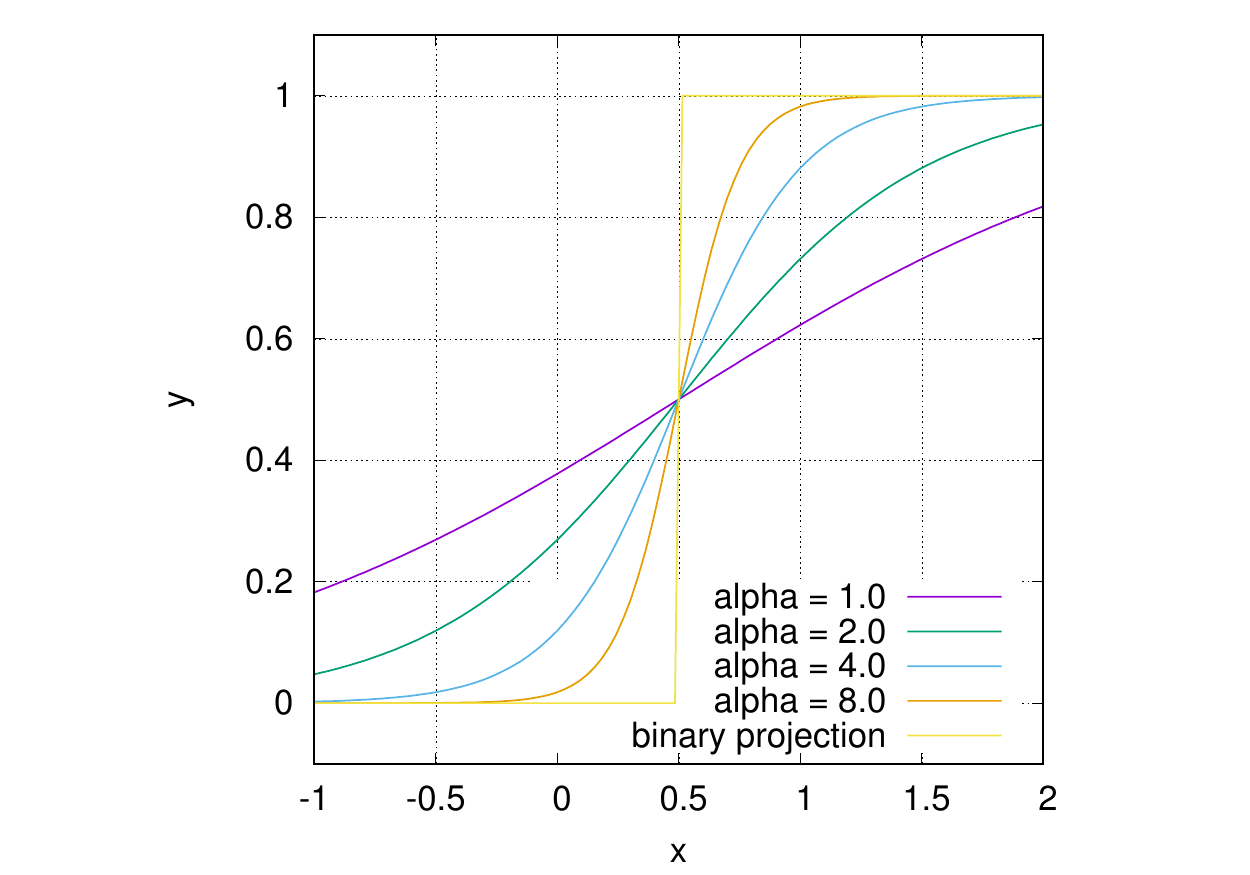}
\end{center}	
\caption{Plots of the shifted sigmoid function $y = \xi(\alpha (x- 0.5))$ for $\alpha = 1.0, 2.0, 4.0, 8.0$. As $\alpha$ gets large, 
the shape of the shifted sigmoid function gradually 
approaches to the binary projection function.}
\label{shifted_sigmoid}
\end{figure}

The projection step is given by 
\begin{equation}
	\bm{s}_{t+1} := \xi \left(\alpha \left(\bm{r}_t - 0.5 \right)  \right),
\end{equation}
where $\xi$ is the sigmoid function defined by
\begin{equation}
		\xi(x) := {1}/{(1 + \exp(-x))}.
\end{equation}
The parameter $\alpha$ controls the softness of the projection.
Precisely speaking, the function $\xi$  is not the projection to 
the binary symbols $\{0, 1\}$. The projection step exploits {\em soft-projection} based on 
the shifted sigmoid function (See Fig. \ref{shifted_sigmoid}) 
because the true projection 
to discrete values results in insufficient convergence behavior in a minimization process.

The main process of the proposed decoding algorithm described later 
is the iterative process executing the gradient step  and
the projection step.

\subsection{Penalty function and objective function}

The penalty function corresponding to the parity constraints is defined by
\begin{equation}
	P(\bm{x}) := \frac 1 2 \sum_{i \in [m]} \sum_{S \in O_i} \left[ \nu \left(1 + \sum_{t \in S} (x_t -1) 	- \sum_{t \in A_i \backslash S} x_t \right) \right]^2
\end{equation}
where the function $\nu$ is the ReLU function defined by
$
	\nu(x) := \max\{0, x \}.
$
This penalty function is a standard penalty function corresponding to the parity polytope ${\cal Q}(H)$
based on the quadratic penalty. 
From this definition of the penalty function $P(\bm{x})$, we immediately have
$
P(\bm{x}) = 0
$
if $\bm{x} \in {\cal Q}(H)$
and 
$
P(\bm{x}) > 0
$
if  $\bm{x} \notin {\cal Q}(H)$.

In the proposed decoding algorithm to be described later, the gradient of the penalty function is needed.
The partial derivative of $P(\bm{x})$ with respect to the variable $x_k$ $(k \in [n])$ is given by
\begin{eqnarray} \nonumber
	\frac{\partial}{\partial x_k} P(\bm{x}) &=& 
	\sum_{i \in [m]} \sum_{S \in O_i} 
	\nu \left(1 + \sum_{t \in S} (x_t -1) 	- \sum_{t \in A_i \backslash S} x_t \right) \\ \label{P_grad}
	&\times& 	\left(\Bbb{I} [k \in S ] - \Bbb{I}[k \in A_i \backslash S] \right).
\end{eqnarray}

As described before, the objective function to be minimized in a decoding process 
is given by
\begin{equation}
	f_\beta(\bm{x}) = \bm{\lambda} \bm{x}^T + \beta P(\bm{x}).	
\end{equation}
The first term of the objective function prefer a point close to the received word.
On the other hand, the second term prefer a point in the parity polytope.
The partial derivative of the objective function with respect to the variable $x_k$ is thus given by
\begin{equation}
	\frac{\partial}{\partial x_k} f_\beta(\bm{x}) 
	= \lambda_k + \beta \frac{\partial}{\partial x_k} P(\bm{x}).
\end{equation}

\subsection{Concise representation of gradient vector}

For the following argument, it is useful to introduce a concise representation of the 
gradient vector

For the odd size family $O_i (i \in [m])$, we prepare a bijection $\phi_i: O_i \rightarrow [2^{|A_i|-1}]$.
For example,  in the case of $A_i  := \{1,2,3\}$, we have
$
O_i := \{\{1\}, \{2\}, \{3\}, \{1, 2, 3\}    \}.
$
A possible choice of $\phi_i$ is as follows:
\[
	\phi_i(\{1\}) = 1,\  \phi_i(\{2\}) = 2,\ 	\phi_i(\{3\}) = 3, \	\phi_i(\{1, 2, 3 \}) = 4. 
\]
Let 
\begin{equation}
	L :=  \sum_{i = 1}^{m} 2^{|A_i|-1},
\end{equation}
which indicates the total number of parity constraints required to define ${\cal Q}(H)$.
The function $\ell(i, S) (i \in [m], S \subset O_i)$ defined by
\begin{equation}
\ell(i, S) := \phi_i(S) +  \sum_{k = 1}^{i-1} 2^{|A_k|-1}
\end{equation}
is a bijection from the set $\{(i, S) : i \in [m], S \subset O_i\}$ to $[L]$.

Let us see a simple example.
Suppose that $H$ is given by
\begin{equation} 
	H = 
	\left(
	\begin{array}{cccccc}
		1 & 1 & 1 & 0 & 0 & 0 \\
		0 & 0 & 1 & 1 & 0 & 0 \\
		0 & 0 & 0 & 1 & 1 & 1 \\
	\end{array}
	\right).
	\label{pm_example}
\end{equation}
A set of bijections $\{\phi_i \}_{i \in [3]}$ defines 
the following $\ell(i, S)$:
\begin{eqnarray} \nonumber
	i = 1, S = \{1 \} &\rightarrow& \ell (i, S) = 1 \\ \nonumber
	i = 1, S = \{2 \} &\rightarrow& \ell (i, S) = 2 \\ \nonumber
	i = 1, S = \{3 \} &\rightarrow& \ell (i, S) = 3 \\ \nonumber
	i = 1, S = \{1, 2, 3 \} &\rightarrow& \ell (i, S) = 4 \\ \nonumber
	i = 2, S = \{3 \} &\rightarrow& \ell (i, S) = 5 \\ \nonumber
	i = 2, S = \{4 \} &\rightarrow& \ell (i, S) = 6 \\ \nonumber
	i = 3, S = \{4 \} &\rightarrow& \ell (i, S) = 7 \\ \nonumber
	i = 3, S = \{5 \} &\rightarrow& \ell (i, S) = 8 \\ \nonumber
	i = 3, S = \{6 \} &\rightarrow& \ell (i, S) = 9 \\ \nonumber
	i = 3, S = \{4, 5, 6 \} &\rightarrow& \ell (i, S) = 10.
\end{eqnarray}

The following matrices $Q$ and $R$ play a key role to derive 
a concise representation of the gradient vector.
The matrix $Q \in \{0, 1\}^{n \times L}$
satisfies 
\[
Q_{i,j} = \left\{
\begin{array}{ll}
1,  &  \mbox{if }  i \in S \mbox{ and } j = \ell(i, S) \\
0, &  \mbox{otherwise} 
\end{array}
\right.
\]
for any $i \in [m]$ and for any $S \subset O_i$.
In a similar way, the matrix $R \in \{0, 1\}^{n \times L}$
satisfies 
\[
R_{i,j} = \left\{
\begin{array}{ll}
1,  &  \mbox{if }  i \in A_i \backslash S \mbox{ and } j = \ell(i, S) \\
0, &  \mbox{otherwise} 
\end{array}
\right.
\]
for any $i \in [m]$ and for any $S \subset O_i$.

We can see that the column order of $Q$ and $R$ depends on 
the choice of the bijections $\{\phi_i \}_{i \in [m]}$, i.e., 
a different choice of $\{\phi_i \}_{i \in [m]}$ yields 
a column permuted version of $Q$ and $R$.
However, in the following argument, the column order does not
cause any influence for gradient computation. We thus can choose any 
set of bijections $\{\phi_i \}_{i \in [m]}$.

Suppose that $H$ is given by (\ref{pm_example}).
A set of bijections $\{\phi_i \}_{i \in [3]}$ is also given.
From the definition of $Q$ and $R$, and $\ell(\cdot, \cdot)$, we have 
\begin{equation}
	Q = 
	\left(
	\begin{array}{cccccccccc}
		1 & 0 & 0 & 1 & 0 & 0 & 0 & 0 & 0 & 0\\
		0 & 1 & 0 & 1 & 0 & 0 & 0 & 0 & 0 & 0\\
		0 & 0 & 1 & 1 & 1 & 0 & 0 & 0 & 0 & 0\\
		0 & 0 & 0 & 0 & 0 & 1 & 1 & 0 & 0 & 1\\
		0 & 0 & 0 & 0 & 0 & 0 & 0 & 1 & 0 & 1\\
		0 & 0 & 0 & 0 & 0 & 0 & 0 & 0 & 1 & 1\\
	\end{array}
	\right),
\end{equation}
\begin{equation}
	R = 
	\left(
	\begin{array}{cccccccccc}
		0 & 1 & 1 & 0 & 0 & 0 & 0 & 0 & 0 & 0\\
		1 & 0 & 1 & 0 & 0 & 0 & 0 & 0 & 0 & 0\\
		1 & 1 & 0 & 0 & 0 & 1 & 0 & 0 & 0 & 0\\
		0 & 0 & 0 & 0 & 1 & 0 & 0 & 1 & 1 & 0\\
		0 & 0 & 0 & 0 & 0 & 0 & 1 & 0 & 1 & 0\\
		0 & 0 & 0 & 0 & 0 & 0 & 1 & 1 & 0 & 0\\
	\end{array}
	\right).
\end{equation}

We are now ready to derive a concise expression of the gradient.
By rewriting (\ref{P_grad}) with the matrices $Q$ and $R$, we have
\begin{equation}
\nabla P(\bm{x}) = \nu (1 + (\bm{x}-1)Q - \bm{x} R) D^T,
\end{equation}
where $D :=  Q - R$.
By using this expression, the gradient vector $\nabla f_\beta(x)$
can be concisely rewritten by
\begin{equation} \label{concise_grad}
\nabla f_\beta(\bm{x}) = \bm{\lambda} + \beta \nu (1 + (\bm{x}-1)Q - \bm{x} R) D^T.
\end{equation}
From this expression, the evaluation of the gradient vector is based on 
the evaluation of the matrix-vector products with sparse matrices $Q, R$, and $D$.
The computational complexity of the gradient vector is to be discussed in the next subsection.

Let us see a simple example of the $3 \times 6$ parity check matrix (\ref{pm_example}).
If $\bm{x}$ is a codeword of $C(H)$, e.g.,  $\bm{x} := (0, 1,1,1,1,0)$, the gradient of the penalty term
$\nabla P(\bm{x})$ becomes the zero vector. This is because the value of the penalty function is constant (i.e., zero)
in the parity polytope. Another example is that 
the center point of the parity polytope $\bm{x} := (1/2, 1/2,1/2,1/2,1/2,1/2)$ gives also the zero gradient.
On the other hand, a non-codeword binary vector
results in a non-zero gradient, e.g., $\nabla P(1,1,1,1,1,0) = (1,1,1,0,0,0)$.
The gradient of the penalty term becomes non-zero for non-codeword binary vector.
This property is useful for decoding processes because a search point repels non-codeword binary vectors 
in gradient descent processes.

\subsection{Trainable Projected Gradient Decoding}
The following decoding algorithm
is based on the projected gradient descent algorithm described in the previous 
subsections.

\noindent
{\underline{Trainable Projected Gradient (TPG) Decoding}}
\begin{itemize}
	\item Input:  received word $\bm{y} \in \mathbb{R}^n$
	\item Output: estimated word $\hat{\bm{c}} \in C(H)$
	\item Parameters:  $t_{max}$: maximum number of the projected gradient descent iterations (inner loop), 
     $r_{max}$: maximum number of restarting (outer loop)
\end{itemize}
\begin{description}
	\item [Step 1] (initialization for restarting) The restarting counter is initialized to $r := 1$.
	\item [Step 2] (random initialization) The initial vector $\bm{s}_1 \in \mathbb{R}^n$ is randomly initialized, i.e.,  each elements in $\bm{s}_1$ is chosen uniformly at random in $\{x \in \mathbb{R} \mid 0 \le x \le 1 \}$. The iteration index is initialized to $t := 1$.
	\item [Step 3] (gradient step) Execute the gradient descent step:
	\begin{equation} \label{pgd_gradient_step}
	\bm{r}_t  := \bm{s}_t - \gamma_t \left( \bm{y} +  \beta_t \nu(1 + (\bm{s}_t - 1)Q - \bm{s}_t R) D^T \right).
	\end{equation}
	\item[Step 4] (projection step) Execute the projection step:
	\begin{equation}	
	\bm{s}_{t+1} := \xi \left(\alpha \left(\bm{r}_t - 0.5 \right)  \right).
	\end{equation}
	\item[Step 5] (parity check) Evaluate a tentative estimate 
	$\hat{\bm{c}} := \theta(\bm{s}_{t+1})$ where the function $\theta$ is the thresholding function 
	defined by
	\begin{equation}
		\theta(x) := 
		\left\{
		\begin{array}{cc}
		0, & x < 0.5, \\
		1, & 	x \ge 0.5.
		\end{array}
		\right.
	\end{equation}
	If $H \hat{\bm{c}} = \bm{0}$ holds, then output $\hat{\bm{c}}$ and exit.
	\item[Step 6] (end of inner loop) If $t < t_{max}$ holds, then $t := t+1$ and go to Step 3.
	\item[Step 7] (end of outer loop) If $r < r_{max}$ holds, then $r := r+1$ and go to Step 2; 
	Otherwise, output $\hat{\bm{c}}$ and quit the process.
\end{description}

The trainable parameters $\{\gamma_t \}_{t = 1}^{t_{max}}$ control the step size 
in the gradient descent step and $\{\beta_t \}_{t = 1}^{t_{max}}$ defines
relative strength of the penalty term. The trainable parameter $\alpha$ controls 
the softness of the soft-projection.
These parameters are adjusted in a training process described later.
In the gradient step, we use the received word $\bm{y}$ instead of the log likelihood ratio vector  
$\bm{\lambda}$ since $\lambda_i \propto y_i $ for $i \in [n]$ under the assumption of the AWGN channel. 
The proportional constant can be considered to be involved in the step size parameter $\gamma_t$.
The parity check in Step 5 helps early termination that may reduce
the expected number of decoding iterations.

The TPG decoding is a double loop algorithm. The inner loop (starting from Step 2 and ending at Step 6)
is a projected gradient descent process such that a search point gradually 
approaches to a candidate codeword as the number of
iterations grows. The outer loop (starting from Step 1 and ending at Step 7) is for executing multiple search processes with
different initial point.  The technique is called {\em restarting}.
The initial search point of TPG decoding $\bm{s}_1$  is randomly chosen in Step 2.
In non-convex optimization, restarting with a random initial point is a basic technique to find 
a better sub-optimal solution.


The most time consuming operation in the TPG decoding is the gradient step (\ref{pgd_gradient_step}).
We here discuss the computational complexity of the gradient step.
In order to simplify the argument, we assume an $(\ell, r)$-regular LDPC code where $\ell$ and $r$ stands for
the column weight and the row weight, respectively.
In the following time complexity analysis, we will focus on the number of multiplications because it 
dominates the time complexity of the algorithm.
Since the number of non-zero elements in $Q$ and $R$ is $m 2^{r - 1}$, the number of 
required multiplications over real numbers for evaluating $(\bm{s}_t-1) Q$ and $\bm{s}_t R$ is $m 2^{r - 1}$.
On the other hand, the multiplication regarding $D^T$ needs $m 2^{r}$ multiplications
because the number of non-zero elements in $Q - R$ is $m 2^{r}$.
In summary, the computation complexity of the TPG decoding is $O(m 2^r)$ per iteration.

\subsection{Training Process}

As we saw in the previous subsection, TPG decoding contains several 
adjustable parameters. It is crucial to train and optimize these parameters
appropriately for achieving reasonable decoding performance.

Let the set of trainable parameters be 
\begin{equation}
\Theta_{t} := \{\alpha,  \{\beta_t \}_{t = 1}^{t},  \{\gamma_t \}_{t = 1}^{t} \} (t \in [t_{max}]).
\end{equation}
Based on a random initial point $\bm{s}_1$ and $\Theta_{t}$,  
we define the function $g_{\bm{s}_1}^{t}: \mathbb{R}^n \rightarrow \mathbb{R}^n$
by $g_{\bm{s}_1}^{t}(\bm{y}) := \bm{s}_{t+1} (t \in [t_{max}])$ where $\bm{s}_{t + 1}$ is given by the recursion:
\begin{eqnarray} \label{g_step}
\bm{r}_t \hspace{-2mm}  &:=& \hspace{-2mm} \bm{s}_t - \gamma_i \left( \bm{y} +  \beta_t \nu(1 + (\bm{s}_t - 1)Q - \bm{s}_t R) D^T \right) \\ \label{p_step}
\bm{s}_{t + 1}\hspace{-2mm}  &:=&\hspace{-2mm}  \xi \left(\alpha \left(\bm{r}_t - 0.5 \right)  \right).
\end{eqnarray}
In other words,  $g_{\bm{s}_1}^{t}(\bm{y})$ represents the search point of a projected gradient descent 
process after $t$ iterations.
In the training process of $\Theta_t$, we use mini-batch based training with a SGD-type parameter update.

Suppose that a mini-batch consists of $K$-triples: 
\begin{equation}
B := \{(\bm{c}_1, \bm{y}_1, \bm{s}_{1,1}), (\bm{c}_2, \bm{y}_2, \bm{s}_{1,2}),\ldots,  (\bm{c}_K, \bm{y}_K, \bm{s}_{1,K}) \}
\end{equation}
which is a randomly generated data set according to the channel model.
The vector $\bm{c}_k \in C(H) (k \in [K])$ is a randomly chosen codeword
and $\bm{y}_k = (1- 2 \bm{c}_k) + \bm{w}_k$   is a corresponding received word where $\bm{w}_k$ is a Gaussian noise vector.
The vector $\bm{s}_{1, k} \in \mathbb R^n (k \in [K])$ is chosen from the $n$-dimensional unit cube uniformly at random,
where these vectors are used as the random initial values.

\begin{figure}
\begin{center}
	\includegraphics[scale=0.4]{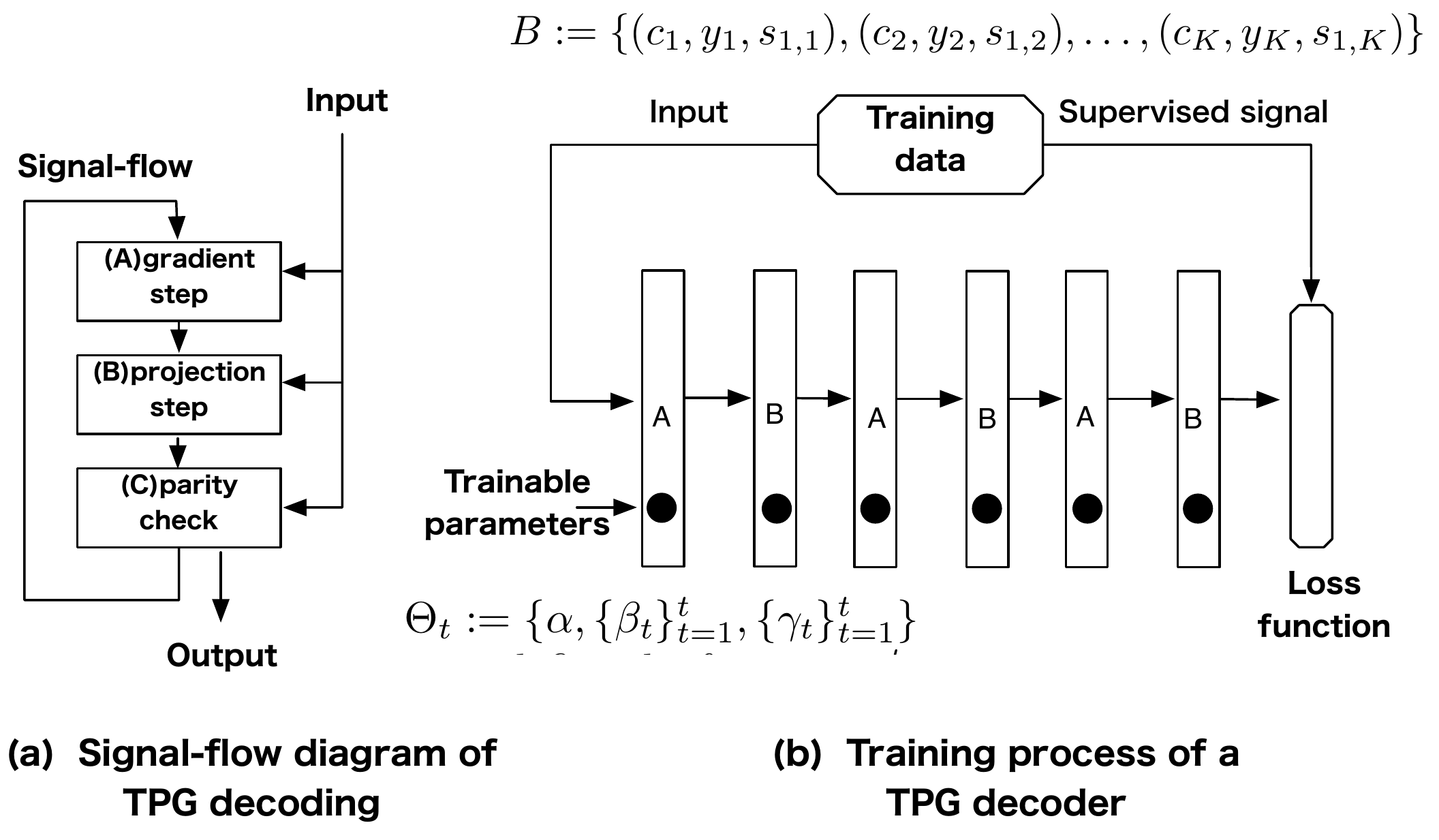}
\end{center}	
\caption{Signal-flow diagram of TPG decoding and training process of a TPG decoder}
\label{sig_flow}
\end{figure}

We exploit a simple squared loss function given by
\begin{equation}
	h^t(\Theta_t) := \frac 1 K \sum_{k =1}^K ||\bm{c}_k - g_{\bm{s}_{1,k}}^{t}(\bm{y}_k)||_2^2
\end{equation}
for a mini-batch $B$. A back propagation process evaluates the gradient $\nabla h^t(\Theta_t)$ and 
it is used for updating the set of parameters as $\Theta_t:= \Theta_t + \Delta_{\Theta_t}$ where 
$\Delta_{\Theta_t}$ is determined by a SGD type algorithm such as AdaDelta, RMSprop, or Adam.
Note that a mini-batch of size $K$ is randomly renewed for each parameter update.
Figure \ref{sig_flow} illustrates the signal-flow diagram of TPG decoding and 
the corresponding unfolded graph used for a training process.

In order to achieve better decoding performance and stable training processes, we exploit 
incremental training such that 
$h^1(\Theta_1), h^2(\Theta_2), \ldots, h^{t_{max}}(\Theta_{t_{max}})$ are sequentially minimized \cite{TISTA}.
The details of the incremental training is as follows. At first,  $\Theta_1$ is trained by minimizing $h^1(\Theta_1)$.
After finishing the training of $\Theta_1$, 
the values of trainable parameters in $\Theta_1$ are copied to 
the corresponding parameters in $\Theta_2$.
In other words, the results of the training for $\Theta_1$ are taken over to $\Theta_2$ as the initial values.
Then,  $\Theta_2$ is trained by minimizing $h^2(\Theta_2)$.
Such processes continue from $\Theta_1$ to $\Theta_{t_{max}}$.
The number of iterations for training $\Theta_i$, which is referred to as a {\em generation}, 
 is fixed to $J$ for all $i \in [t_{max}]$.

In this work, the training process was implemented by PyTorch \cite{PyTorch}.

\section{Experimental Results}

In this section, we will show several experimental results indicating the behavior and 
the decoding performance of the TPG decoding.

\subsection{Behavior of TPG decoding}

We trained a TPG decoder with 
a $(3,6)$-regular LDPC code of $n = 204, m = 102$.
Several hyper parameters assumed in the training process are 
as follows: The maximum number of iterations is set to $t_{max} = 25$.
The number of parameter updates for a generation is $J = 500$ and 
the mini-batch size is set to $K = 50$.  We employed Adam optimizer \cite{Adam}
with learning rate $0.005$ for the parameter updates. In a training process, 
the SNR of channel is fixed to $SNR = 4.0$ dB.

\begin{figure}
\begin{center}
	\includegraphics[scale=0.7]{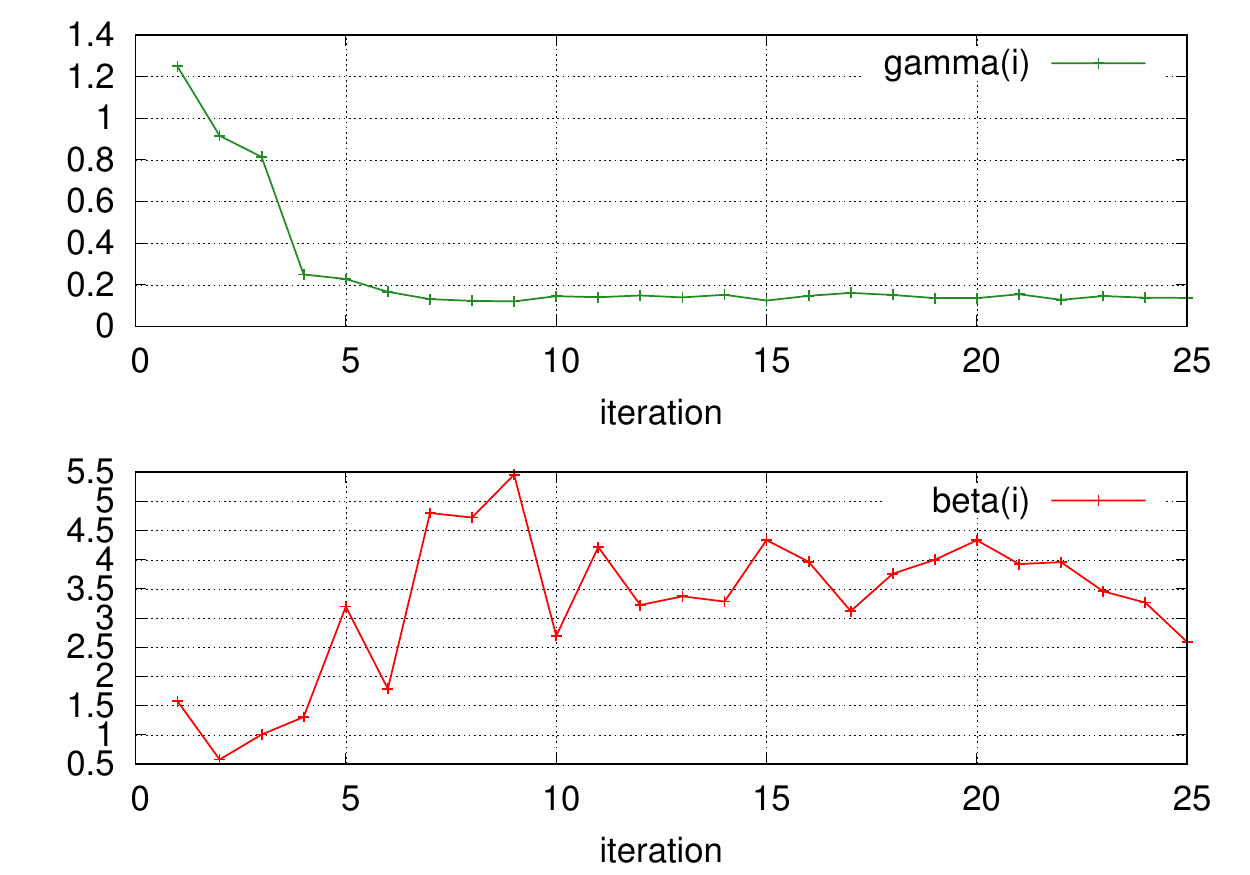}
\end{center}	
\caption{Plots of trained parameters $\{\gamma_t\}_{t=1}^{25}$ and $\{\beta_t\}_{t=1}^{25}$ ($n = 204, m  = 102$)}
\label{train}
\end{figure}

Figure \ref{train} indicates the result of the training, i.e., 
trained parameters $\{\gamma_t\}_{t=1}^{25}$ and $\{\beta_t\}_{t=1}^{25}$.
At the first iteration, the step size parameter $\gamma_t$ takes the value around $1.2$ and the value gradually decreases to 
the values around $0.2$. On the other hand, the penalty term constant $\beta_i$ starts from 
the small value around $1$ and increases to the values around $5.5$ at the 9-th round.
The softness parameter is a shared trainable variable for all the rounds takes the value $\alpha = 8.05$.

\begin{figure}
\begin{center}
	\includegraphics[scale=0.7]{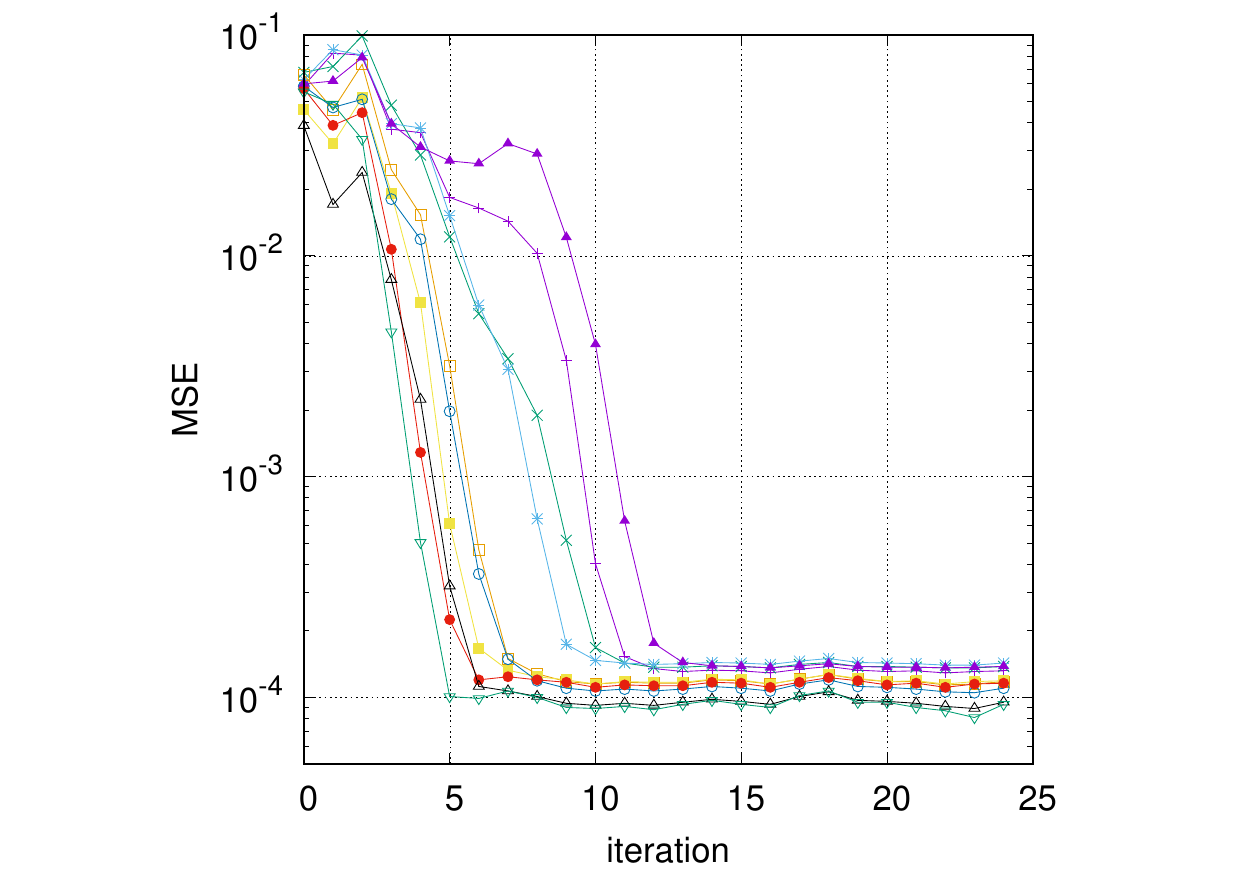}
\end{center}	
\caption{Plots of the trajectories of normalized squared error for 10 trials for a fixed 
received word ($n = 204, m  = 102$, $SNR = 4.0$ (dB))}
\label{exp3}
\end{figure}

In a decoding process of TPG decoding, we expect that the search point
approaches to the transmitted codeword.
In order to observe the behavior of a TPG process based on the recursive formula (\ref{g_step}) (\ref{p_step}), 
we show the trajectories of the normalized squared error in Fig.~\ref{exp3}.
The normalized squared error is defined by $(1/n)||\bm{s}_t - \bm{c}^*||_2^2$ where $\bm{c}^*$ is the 
transmitted codeword.
Figure \ref{exp3} includes the trajectories of 10 trials with random initial values.
A received word $\bm{y} := (1 - 2 \bm{c}^*) + w$ is fixed during the experiment.
The code is the $(3,6)$-regular LDPC code  with $n = 204$ and $m = 102$ and 
the trainable parameters $\{\gamma_t\}_{t=1}^{25}$ and $\{\beta_t\}_{t=1}^{25}$ are set to the 
values in Fig.~\ref{train} and $\alpha$ is set to $8.05$ according to the above training result.
The noise variance is corresponding to $4.0$ (dB).

From Fig.~\ref{exp3}, we can observe that each curve indicates rapid decrease of the normalized squared error 
(around 10 rounds for convergence)
and it means that a search point $\bm{s}_t$ actually approaches to the transmitted word 
in the recursive evaluation of (\ref{g_step}) (\ref{p_step}).
With several iterations (5 to 15), the normalized squared error 
gets to the value around $10^{-4}$. This results implies that the penalty function representing 
the parity constraints are effective to direct the search point towards the transmitted word
and that trained parameters provide intended behavior in minimization processes. 
Another observation obtained from Fig.~\ref{exp3} is that search point trajectories are different from 
each other and that are dependent on the initial value. 
The idea of restarting is based on the expectation that random initial values provide random outcomes.
The experimental results support this expectation.

\subsection{BER performances}

Several hyper parameters assumed in the training process are 
as follows: 
The number of parameter updates for a generation is $J = 500$ and 
the mini-batch size is set to $K = 50$.  We employed Adam optimizer \cite{Adam}
with learning rate $0.005$ for the parameter updates. In a training process, 
the SNR of channel is fixed to $SNR = 4.0$ dB.

\begin{figure}
\begin{center}
	\includegraphics[scale=0.7]{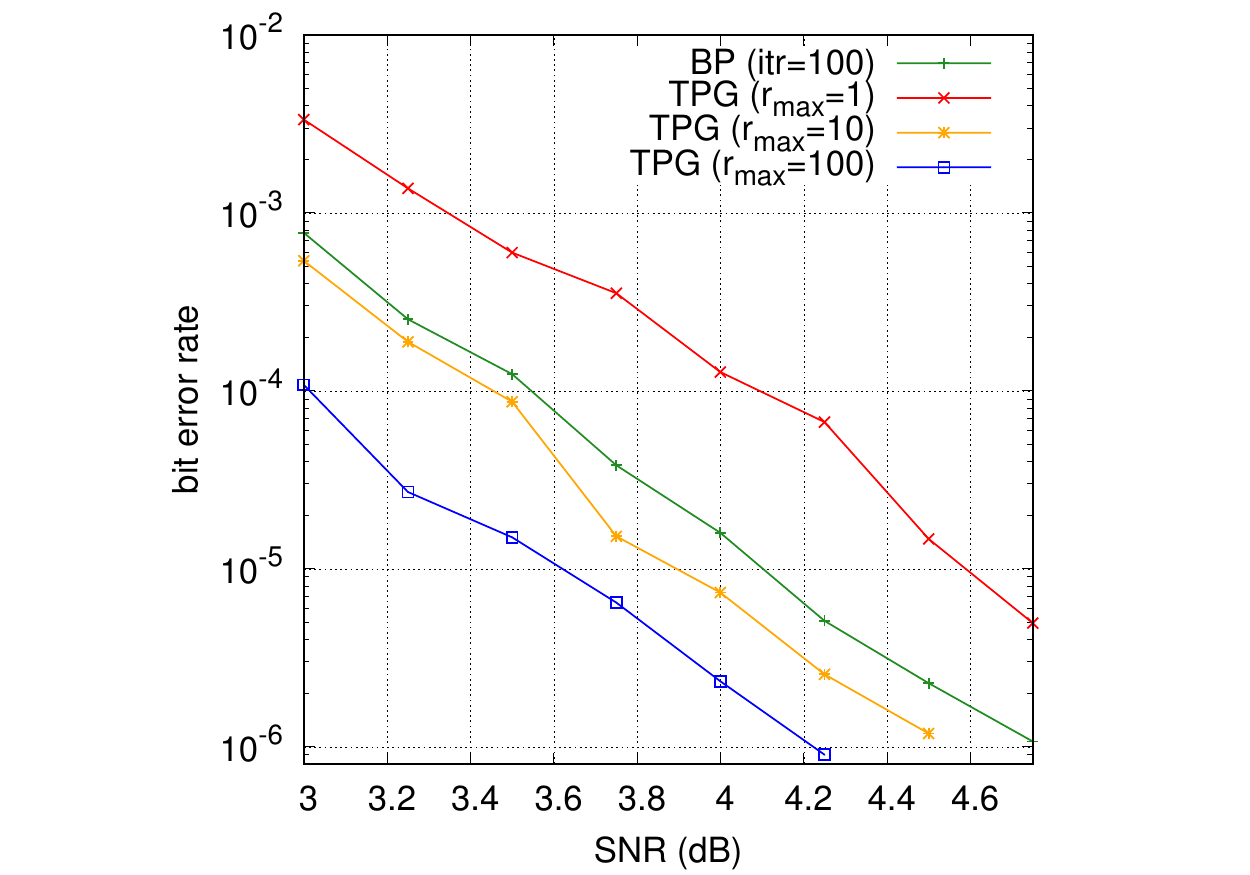}
\end{center}	
\caption{BER performance of the TPG decoding for (3.6)-regular LDPC code ($n = 204, m  = 102$). 
Parameters: $t_{max} = 100, K = 50, J = 500$, 
training $SNR = 4.0$ (dB), Adam optimizer with learning rate $0.005$}
\label{BERperformance204}
\end{figure}

\begin{figure}
\begin{center}
	\includegraphics[scale=0.7]{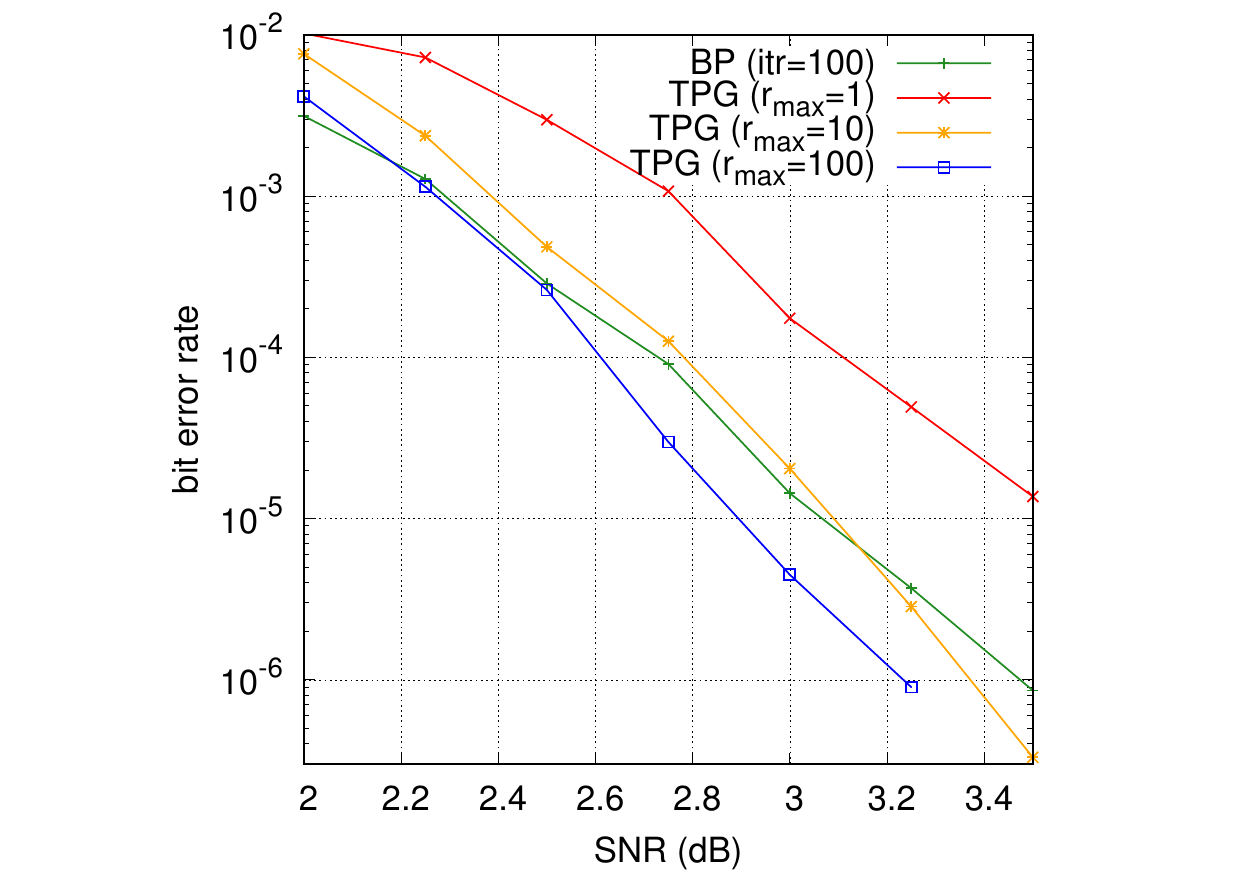}
\end{center}	
\caption{BER performance of the TPG decoding for (3,6)-regular LDDP code ($n = 504, m  = 252$).
Parameters: $t_{max} = 100, K = 50, J = 500$, 
training $SNR = 2.5$ (dB), Adam optimizer with learning rate $0.001$
}
\label{BERperformance504}
\end{figure}

The decoding performances of TPG decoding 
for the rate $1/2$ (3,6)-regular LDPC code with $n = 204$
are shown in Fig. \ref{BERperformance204}. 
Figure \ref{BERperformance204} includes the BER curves of 
TPG decoding with $r_{max} = 1, 10, 100$.
As the baseline performance, the BER curve of the belief propagation (BP) decoding where
the maximum number of iterations is set to 100.
The BER performance of TPG decoding  ($r_{max} = 1$) is inferior to 
that of BP.  On the other hand, we can observe that restarting significantly 
improves the decoding performance of the proposed algorithm.  In the case of $r_{max} = 10$,
the proposed algorithm shows around $0.2$ dB gain over the BP at  BER$=10^{-5}$
In the case of $r_{max} = 100$,  the proposed algorithm outperforms the BP
and yields impressive improvement in BER performance.
For example, it achieves $0.5$ dB gain at BER $= 10^{-5}$.
These results  indicate that restarting works considerably well as we expected.
This means that we can control trade-off between decoding complexity and the decoding 
performance in a flexible way.
Figure \ref{BERperformance504} shows
the BER curves of TPG decoding 
for the rate $1/2$ (3,6)-regular LDPC code with $n = 504$.
We can observe that the proposed algorithm again provides superior BER performance 
in the high SNR regime.

\begin{figure}
\begin{center}
	\includegraphics[scale=0.7]{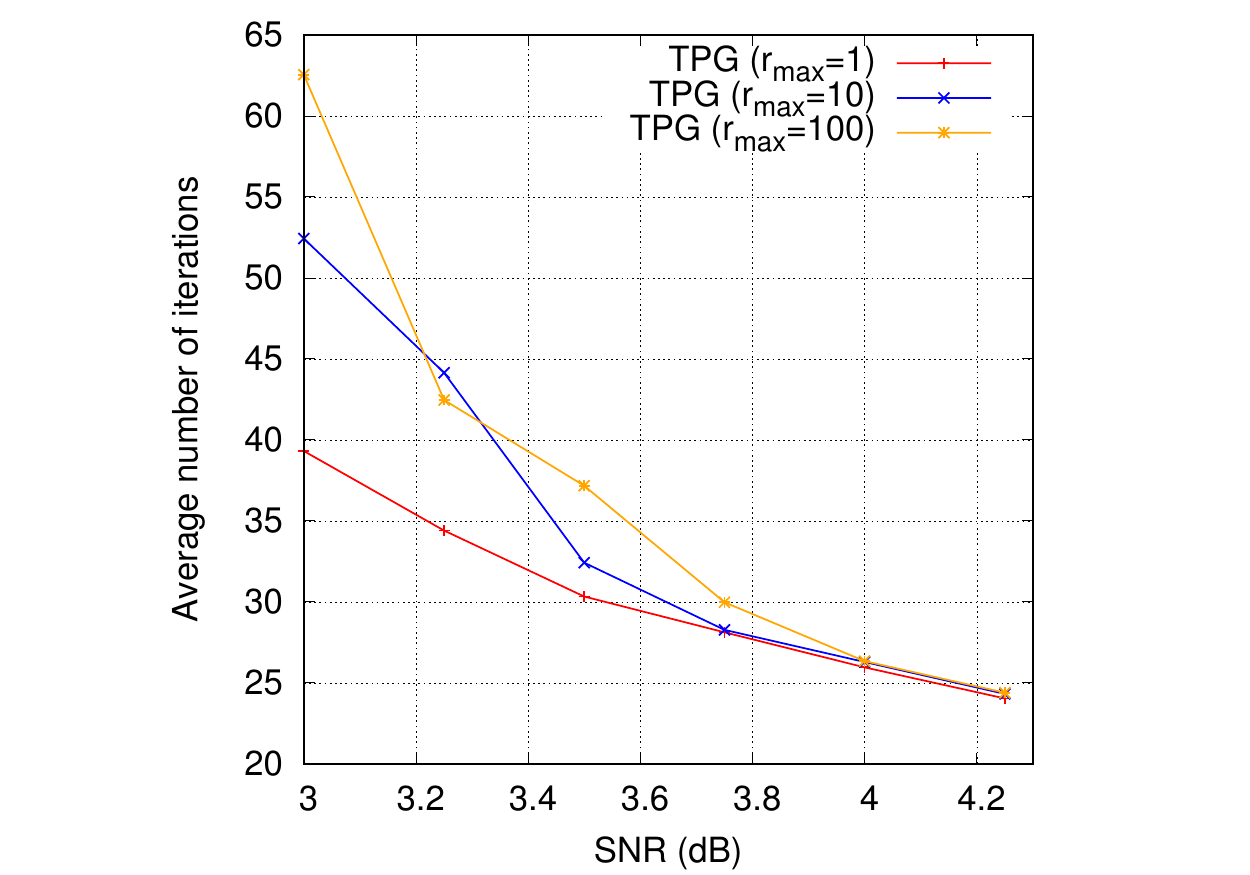}
\end{center}	
\caption{Average number of iterations of the TPG decoding for (3,6)-regular LDDP code ($n = 204, m  = 102$).
Parameters: $t_{max} = 100$.}
\label{Ave_itr}
\end{figure}

The average time complexity of the proposed decoding algorithm is closely related to 
the average number of iterations in the TPG decoding processes.
Early stopping by the parity check (Step 5) reduces the number of iterations.
The number of iterations means the number of execution of Step 3 (gradient step) for 
a given received word.  
The average number of iterations for (3,6)-regular LDDP code ($n = 204, m  = 102$) are plotted in Fig. \ref{Ave_itr}.
When SNR is 3.75 dB, the average number of iterations is around 30 for all the cases ($r_{max} = 1, 10, 100$).

\section{Concluding summary}

In this paper, we present a novel decoding algorithm for LDPC codes, which is based on 
a non-convex optimization algorithm. The main processes in the proposed algorithm are the 
gradient and projection steps that have intrinsic massive parallelism that fits 
forthcoming deep neural network-oriented hardware architectures.
Some of internal 
parameters can be optimized with data-driven training process with back propagation and
a stochastic gradient type algorithm. Although we focus on the AWGN channels in this paper,
we can apply TPG decoding for other channels such as linear vector channels just by replacing 
the objective function.

\section*{Acknowledgement}
This work was partly supported by JSPS Grant-in-Aid for Scientific Research (B) 
Grant Number 16H02878.



\end{document}